\documentclass[aps,pra,reprint,superscriptaddress, longbibliography,noeprint]{revtex4-2}
\usepackage{lipsum}

\usepackage{siunitx}
\DeclareSIUnit\gauss{G}
\usepackage{amsmath,bm}
\usepackage{dsfont}
\usepackage{amsbsy}
\usepackage{amssymb}
\usepackage{mathptmx, textcomp}
\usepackage{color}
\usepackage{braket}
\usepackage{graphicx}
\usepackage{textcomp}
\usepackage{notes2bib}
\usepackage{textcomp}
\usepackage{mwe}
\usepackage{siunitx}
\usepackage{filecontents}
\usepackage{soul}
\definecolor{bl}{rgb}{0, .1, .6}
\usepackage[colorlinks=true, citecolor = bl, linkcolor = bl, urlcolor=bl, pdfborder={0 0 0}]{hyperref}

\newcommand{\as}{\alpha_{S}}
\newcommand{\av}{\alpha_{V}}
\newcommand{\at}{\alpha_{T}}
\newcommand{\sys}{_{\rm sys}}
\newcommand{\stat}{_{\rm stat}}
\begin{document}
	\title{Anisotropic polarizability of Dy at 532\,nm on the intercombination transition}

\author{Damien Bloch}\email{damien.bloch@institutoptique.fr}
\author{Britton Hofer}
\affiliation{Universit\'e Paris-Saclay, Institut d'Optique Graduate School, CNRS, 
Laboratoire Charles Fabry, 91127, Palaiseau, France}
\author{Sam R.~Cohen}\altaffiliation{Present address: Department of Physics, Stanford University, Stanford, California 94305, USA}
\author{Maxence Lepers}
\affiliation{Laboratoire Interdisciplinaire Carnot de Bourgogne, CNRS, Université de Bourgogne, 21078 Dijon, France}
\author{Antoine Browaeys}
\author{Igor Ferrier-Barbut}\email{igor.ferrier-barbut@institutoptique.fr}
\affiliation{Universit\'e Paris-Saclay, Institut d'Optique Graduate School, CNRS, 
Laboratoire Charles Fabry, 91127, Palaiseau, France}
\begin{abstract}
    We report experimental measurements of the dynamical polarizability of dysprosium, at a wavelength of \SI{532}{\nano\meter}. We measure all three components (scalar, vector, tensor) of the anisotropic polarizability for the ground and the excited manifolds of the intercombination transition of Dy at \SI{626}{\nano\meter}. The apparatus on which the measurements are performed is first presented. We obtain with this setup imaging of single Dy atoms with fidelity above 99\,\% and losses below 2.5\,\% induced by imaging. We then describe the methods used to extract the polarizability. In particular, we combine a measurement of trap frequency and trap depth on single atoms in optical tweezers, allowing us to obtain a measurement of the ground state polarizability free of errors in trap waist calibration. The obtained values give a magic condition between two Zeeman states in the ground and excited manifolds, which was used to image single atoms in optical tweezer arrays. The scalar polarizability of the ground state is in disagreement with theoretical expectations, calling for future investigations to resolve the discrepancy.
\end{abstract}
\maketitle

\section{Introduction}
In recent years, cold-atom experiments using lanthanide atoms have made possible the exploration of a wide variety of problems in quantum physics thanks to their large spin and high magnetic moment \cite{Chomaz2023}. 
New experimental setups are now required for the application of state-of-the-art atomic physics tools such as optical tweezer arrays \cite{Barredo2016,Endres2016,Kim2016} or quantum gas microscopy \cite{Bakr2009,Sherson2010} to lanthanides. The latest was recently adapted \cite{su2023,Kirilov2023}, and we reported the trapping of single dysprosium atoms in optical tweezer arrays in \cite{Bloch2023}.

Here we present the apparatus on which single Dy atoms in optical tweezer arrays are produced.
It is based on a three-dimensional magneto-optical trap (MOT) inside a glass cell loaded by a two-dimensional MOT (2D MOT), similar to the methods reported in ref.\,\cite{jin2023}. 
Narrow-line imaging in optical tweezers is strongly impacted by the trap light-shift and hence requires knowledge of the atomic polarizability.
We measure the anisotropic polarizability at a trapping wavelength of \SI{532}{\nano\meter} for the ground and excited manifolds of the intercombination line of Dy at \SI{626}{\nano\meter}. 
From the measured values we identify a magic light polarization for which the lowest Zeeman states of each manifold have the same polarizability, which allows for single atom imaging in the optical tweezers.
We finally compare the experimental values to the theoretical predictions, finding a disagreement in the ground state in particular. This highlights the need for future research to understand the value of the polarizability of Dy at \SI{532}{\nano\meter}.


\section{Experimental setup}



For the first laser cooling stages of the experiment, we use the broad blue transition of $^{162}$Dy between the manifolds $G=4f^{10}6s^2\;^5I_8$ and $F=4f^{10}(^5I_8)6s6p(^1P^o_1)\;(8,1)^o_9$ (see Fig. \ref{fig:apparatus}a). 
Its wavelength is $\lambda=\SI{421}{\nano\meter}$ and its linewidth is $\Gamma_{421}=2\pi \times \SI{32}{\mega\hertz}$. We also make use of the narrower intercombination line at $\lambda=\SI{626}{\nano\meter}$ between $G$ and $E=4f^{10}(^5I_8)6s6p(^3P^o_1)\;(8,1)^o_9$ with linewidth $\Gamma_{626}=2\pi\times\SI{135}{\kilo\hertz}$. This transition is used to realise a 3D MOT and to image the atoms in the optical tweezers. The optical tweezers themselves are realised with far-detuned light at \SI{532}{\nano\meter}.

\subsection{Apparatus overview}\label{subsec:apparatus-description}

\begin{figure*}
    \includegraphics{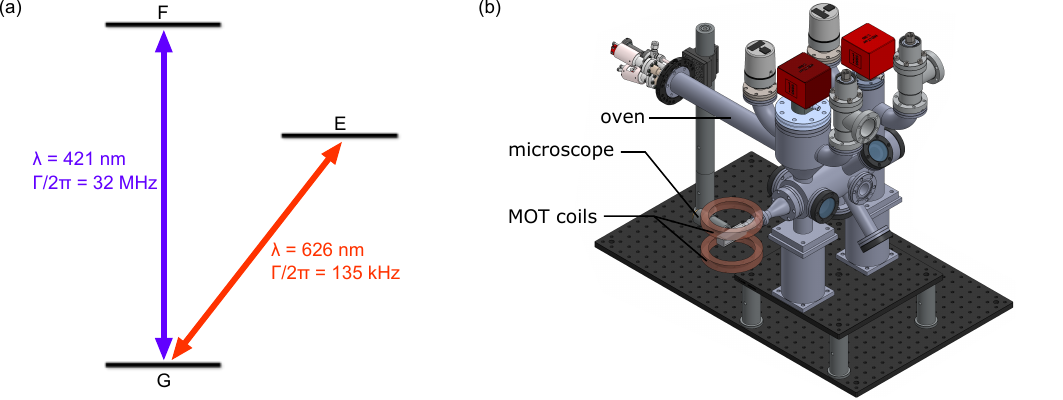}
    \caption{
    (a) Optical transitions of Dy used in this work. 
    (b) Sketch of the vacuum system.
    }
        \label{fig:apparatus}
\end{figure*}

To trap and image single atoms, the combination of a glass cell and a microscope objective has emerged as a versatile solution offering high optical access and compactness \cite{Endres2016,Norcia2018,Cooper2018,Saskin2019}. Such a platform is desirable for lanthanide atoms to perform single atom trapping and imaging in tweezers. One experimental challenge faced in adapting these setups to lanthanides such as Er or Dy is that any surface in line of sight of the oven will be coated by the high atomic flux due to their very high melting point. 
This prevents placing a glass surface or glass cell in fronts of the atomic jet of the oven. 
In order to directly load a glass cell, the solution we implement here is a 2D MOT of Dy similar to the one reported in ref.~\cite{jin2023}, which serves as a continuous source of cold Dy atoms pushed towards the glass cell \footnote{It is also possible to place a \SI{45}{\degree} mirror inside of the glass cell \cite{Kirilov2023}, but this poses stringent requirements to the glass cell making.}. 
This alleviates the need for optical transport and so ensures high experiment repetition rates ($>\SI{1}{\hertz}$).
strontium \cite{Nosske2017,Barbiero2020}. It consists in two parts separated by a differential pumping section (see Fig. \ref{fig:apparatus}b).
The first section is a stainless steel chamber in which a high temperature oven loads a 2D MOT of Dy. 
The oven (Createc DFC-40-25-WK-2) is made of an effusion cell containing the solid Dy and a hotter lip near the front of the oven to prevent Dy deposit from clogging the output. 
The effusion cell is loaded with \SI{70}{\gram} of Dy and heated up to \SI{900}{\celsius} during normal operation, while the hot lip is held at \SI{1200}{\celsius}. 
The chamber is designed such that the output of the oven is only \SI{7}{\centi\meter} away from the center of the 2D MOT, in order to mitigate the effect of the flux divergence at the output of the oven. 
The 2D MOT chamber is made of a main cylinder with 8 radially pointing flanges welded to provide a connection for the oven and access for the 2D MOT beams, as well as additional optical access to image the 2D MOT.
A near-resonant beam on the \SI{421}{\nano\meter} transition is also sent through a viewport to push the atoms from the 2D MOT through the differential pumping section and towards the glass cell in which they are trapped in a 3D MOT. 
The science cell (ColdQuanta $10 \times 13 \times 60$ mm) is surrounded by a pair of anti-Helmholtz coils that produces the quadrupole magnetic gradient for the 3D MOT. 
In addition, three pairs of Helmholtz coils are used to control the homogeneous offset field in the cell. 
The 3D MOT is used to load atoms in optical tweezers, which are made by focusing light at 532 nm through a 0.5 NA microscope objective (Mitutoyo G Plan APO 50x), which we also use to capture single atom fluorescence \cite{Bloch2023}. 
We review below the main stages of the experiment.


\subsection{2D MOT}\label{subsec:2d-mot}

\begin{figure}
    \includegraphics[width=\columnwidth]{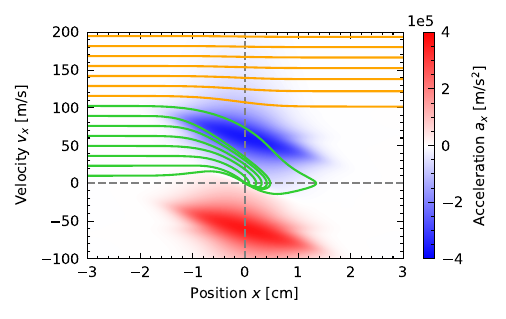}
    \caption{
    Simulated phase space trajectories in the 2D MOT in the plane $x, z=0$.
    Green lines are trajectories of atoms with incoming velocity below the capture velocity of the 2D MOT.
    Orange lines are trajectories of atoms that are too fast to be captured by the 2D MOT and just fly through it.
    The red and blue color map indicates the acceleration experienced by the atoms due to the 2D MOT light.
    }
    \label{fig:2d-mot-trajectories}
\end{figure}

The velocity of the atoms coming out of the oven is about \SI{400}{\meter\per\second}.
To capture the largest fraction of these atoms, we use a 2D MOT on the broad \SI{421}{\nano\meter} transition.
The cooling light is made of a single beam passing in a bow-tie configuration \cite{jin2023}.
The Gaussian beam waist is \SI{8}{\milli\meter} and its power is \SI{300}{\milli\watt} ($I = 5 \times I_{\rm sat}$). It is detuned by $\Delta=-3.2\times \Gamma_{421}$ from the transition. The magnetic field of the 2D MOT is generated with 4 stacks of 5 permanent magnets placed on the corners of a rectangle with a width of 7 cm and a height of 9 cm in the XZ plane. 
These magnets generate a theoretical magnetic gradient of 23 G/cm along the Y and Z directions.
With these parameters, the capture velocity of the 2D MOT is on the order of \SI{100}{\meter\per\second}.
This is visible in figure \ref{fig:2d-mot-trajectories} that shows simulated trajectories of atoms coming out of the oven and crossing the 2D MOT region. 
The simulations are performed using the PyLCP package \cite{Eckel2022}.

To load the 3D MOT in the glass cell, the atoms are pushed in the non-confining direction of the 2D MOT by a near-resonant beam on the \SI{421}{\nano\meter} transition. The glass cell is \SI{30}{\centi\meter} away from the 2D MOT, and the 3D MOT beam radius is about \SI{5}{\milli\meter}. To prevent atoms from dropping under gravity below the 3D MOT beams, they must be accelerated to a velocity $v\gtrsim\SI{15}{\meter\per\second}$.
The push beam has a waist of $w_{\rm push}=\SI{1.2}{\milli\meter}$, a detuning of $\Delta_{\rm push}=-1.9\times\Gamma_{421}$ and an intensity of \SI{0.7}{\milli\watt} ($I_{\rm push}=0.55 \,I_{\rm sat}$).
With these parameters, the velocity reached by the atoms is close to \SI{20}{\meter\per\second}.
To prevent the radiation pressure of this push beam to disturb the 3D MOT in the glass cell, we angle it by about \SI{5}{\degree} so that it does not pass through the differential pumping section and thus does not reach the cell \cite{Wodey2021}.

\subsection{Core-shell MOT}

In the glass cell, we realize a 3D MOT on the \SI{626}{\nano\meter} transition whose relatively narrow linewidth of \SI{135}{\kilo\hertz} results in a low Doppler temperature $T_{\rm D}\simeq\SI{3.2}{\micro\kelvin}$ \cite{Maier2014,Dreon2017}. 
However this linewidth comes also at the cost of a low capture velocity. The waist of the MOT beams is limited to about $w\simeq\SI{1}{\centi\meter}$ by the size of the cell. This yields a capture velocity $v_c \sim \sqrt{\hbar k\Gamma w/m}\lesssim \SI{10}{\meter\per\second}$, lower than the minimum velocity for the atoms to reach the cell. Modulating the MOT frequency \cite{Maier2014,Dreon2017} or using very high light intensity \cite{jin2023} cannot bridge this gap.
To circumvent this limitation and efficiently capture atoms in the 3D MOT, we realise a core-shell MOT \cite{Lee2015} on the \SI{421}{\nano\meter} and \SI{626}{\nano\meter} transitions.
It consists of a hollow outer shell of \SI{421}{\nano\meter} light with a core of \SI{626}{\nano\meter} light. 
Atoms coming from the 2D MOT are efficiently slowed down by the \SI{421}{\nano\meter} shell and can then accumulate in the \SI{626}{\nano\meter} core. The advantage of this configuration is that we benefit from the high capture velocity afforded by the large scattering rate of the blue transition, while avoiding losses due to light-induced collisions and pumping to metastable states \cite{Youn2010,Hoschele2023}.

To realise this core-shell MOT, we overlap on the horizontal \SI{626}{\nano\meter} MOT beams two retro-reflected beams at \SI{421}{\nano\meter} with waists of $w_{421}=\SI{1.5}{\centi\meter}$ and whose centers are blocked by masks of diameter \SI{5}{\milli\meter}. 
A blue beam on the vertical direction is not necessary since the vertical velocity of the atoms is low. 
We find an optimal magnetic gradient for this bi-color MOT to be $B'_{cs}=\SI{5}{\gauss\per\centi\meter}$.
The detuning for the \SI{626}{\nano\meter} MOT during the loading stage is $\Delta_{626}=-40\times \Gamma_{626}$ and the intensity of each beam is $I_{626} = 100 \times I_{\rm sat, 626}$.
The detuning of the \SI{421}{\nano\meter} shell is $\Delta_{421}=-1.3\times \Gamma_{421}$, with a power of about \SI{5}{\milli\watt} per beam (including the blocked central part). 
We observe about a hundredfold increase in atom number when the core of this MOT is not exposed to \SI{421}{\nano\meter} light, demonstrating the reduction of losses by blocking the central part of the blue beams.
Once the 3D MOT is loaded, we abruptly turn off the 2D MOT beams, the push beam and the 3D MOT capture shell to stop the loading. We then slowly ramp the parameters of the \SI{626}{\nano\meter} MOT over \SI{80}{\milli\second} to a gradient of \SI{1.7}{\gauss\per\centi\meter}, a detuning of $-7.4\,\Gamma_{626}$ and a light intensity of $I=3\, I_{\rm sat, 626}$.
After this step, the atoms are polarized in the loweest Zeeman state $|G, J=8, m_J=-8\rangle$ \cite{Maier2014,Dreon2017}.

\subsection{Optical tweezers}

\begin{figure}
    \includegraphics[width=\columnwidth]{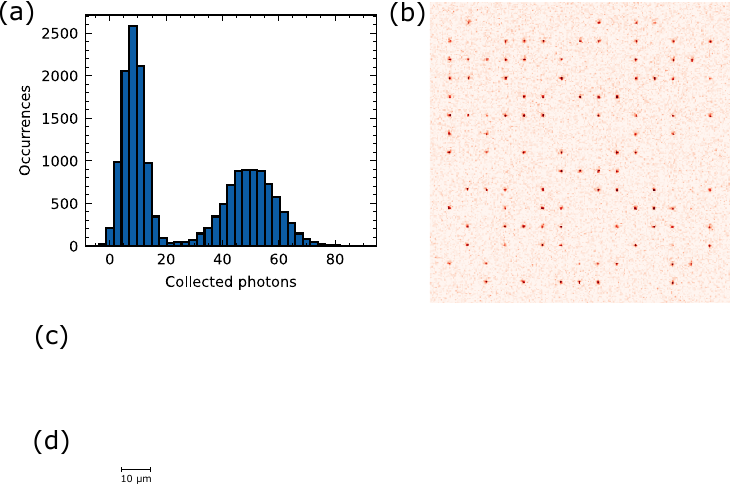}
    \caption{
    (a) Histogram of the collected fluorescence for a given traps over many realisation of the experiment.
    (b) Single shot picture of a 15x15 array of traps randomly loaded. (c, d) Single shot picture of an array of 75 traps before and after sorting the traps.}
    \label{fig:single_atoms}
\end{figure}

In the next step of the experiment we load single Dy atoms in optical tweezers. The loading procedure is detailed in \cite{Bloch2023}. 
Briefly, the tweezer array is generated at \SI{532}{\nano\meter}, using an acousto-optic deflector (AOD) to generate a 1D or 2D array of tweezers, focused by a 0.5-numerical aperture microscope objective. 
The tweezers are loaded from the MOT as described in \cite{Bloch2023}. They are imaged with a single non retro-reflected beam at \SI{626}{\nano\meter} with an intensity $I=0.5 \, I_{\rm sat, 626}$ and a detuning $\Delta=-1.0 \, \Gamma_{626}$. 
Figure \ref{fig:single_atoms}(a) shows a histogram of the collected photons for a given trap over many realisations of the experiment, with the typical bimodal distribution illustrating the random loading of each tweezer.
Having improved the imaging setup with respect to \cite{Bloch2023}, we now obtain a $2.5\,\%$ loss probability after taking a $\SI{30}{\milli\second}$-long image, and an imaging fidelity above $99\,\%$. 
As shown in \cite{Bloch2023}, the losses are mainly due to two-photons events during which an atom in the excited state of the \SI{626}{\nano\meter} transition can absorb a trap photon at \SI{532}{\nano\meter} and decay to highly excited metastable states. 
Since these states have a lifetime of several seconds, the atom will not be detected in the following pictures, and is registered as a loss. The same mechanism was observed for Yb \cite{Saskin2019}.

Representative images of a randomly loaded 2D and 1D array are shown in figures \ref{fig:single_atoms}(b, c).
Figure \ref{fig:single_atoms}(d) shows a typical example of a picture obtained after rearranging the tweezers in 1D using the technique presented in \cite{Endres2016}.

A crucial ingredient for the loading and imaging of single atoms using a cooling transition with a linewidth on the order of \SI{100}{\kilo\hertz} is to obtain a \emph{magic} trapping condition \cite{Saskin2019,Ma2022}. 
The differential light shift between the ground and excited state  can be of the order of several megahertz in optical tweezers with a few hundreds of microkelvin depth.
This is large compared to the linewidth $\Gamma_{626}=2\pi\times 135$~kHz of the intercombination transition at \SI{626}{\nano\meter}. Therefore, this differential shift significantly alters the effect of resonant light on the atoms. It was found for the intercombination line of Yb with a similar linewidth (\SI{182}{\kilo \hertz}), that the magic condition (zero differential light shift) is necessary for single atom imaging \cite{Saskin2019,Jenkins2022}, and we experimentally found the same to be true for Dy.
Finding this magic ellipticity requires the knowledge of the scalar, tensor and vector polarizabilities of the ground and excited state manifolds of the intercombination line. 
We explain in the next section the experimental measurements which allowed us to extract these 6 polarizabilites at \SI{532}{\nano\meter}. From these values we have found a magic ellipticity of the tweezer light for which the ground state $\ket g=\ket{G,J=8,m_J=-8}$ and the excited state $\ket e=\ket{E,J=9,m_J=-9}$ have equal total polarizability.

\section{Polarizabilities at 532 \lowercase{nm}}\label{section:polarizabilities}
\newcommand{\NA}{\text{NA}}

Lanthanides are known to feature a highly anisotropic light-matter interaction \cite{Kotochigova2011}. 
For atom trapping with dipole traps, this anisotropy manifests itself in a strong dependence on light polarization via the vector and tensor components as was observed for instance for Er or Dy \cite{Lepers2014,Kao17,Becher2018,Chalopin2018a,Kreyer2021}. Optical traps for Dy are often in the near infrared at 1070 nm or 1064 nm \cite{Ravensbergen2018}. 
For a trap light at $\lambda_{\rm trap}=\SI{532}{\nano\meter}$, it is known that the ground state of Dy has a positive polarizability so that atoms can be trapped \cite{Maier2015b,Schmitt2017}. 
However, the exact value of the polarizability was not precisely measured at this wavelength, neither for the ground state nor for the excited state of the intercombination transition. 



The effect of the anisotropic polarizability is well described by a light-shift hamiltonian $\hat{H}_{ls}$ that acts separately on each (hyper)fine structure manifold $\hat{H}_{ls}=\hat{H}_S+\hat{H}_V+\hat{H}_T$ that depends on the scalar vector and tensor polarizabilities $\as$, $\av$ and $\at$ as follows \cite{LeKien2013,Dupont1972}:
\begin{equation}
    \begin{split}
    \hat{H}_S&=-\as \frac{I}{2c \epsilon_0}\\
    \hat{H}_V&=-i \av \frac{I}{2c \epsilon_0} (\bm{\epsilon^*}\times\bm{\epsilon})\cdot \frac{\hat{\bm{J}}}{2J}\\
    \hat{H}_T&=-\at \frac{I}{2c \epsilon_0}\frac{3\{\bm{\epsilon} \cdot \hat{\bm{J}}, \bm{\epsilon}^* \cdot \hat{\bm{J}}\}-2\hat{\bm{J}}^2}{2J(2J-1)}
        \end{split}
\end{equation}
where $I$ is the light intensity, $\bm{\epsilon}$ is the complex polarization of the electric field, $\hat{\bm J}$ the atomic spin and $\{\hat{A}, \hat{B}\}=\hat{A}\hat{B} + \hat{B}\hat{A}$ is the anti-commutator between two operators $\hat{A}$ and $\hat{B}$. 
In the presence of a magnetic field as is the case for our experiments, the total hamiltonian can be expressed as:
\[\hat{H}_{\rm tot}=\hat{H}_0 + \hat{H}_{ls}\]
where $\hat{H}_0=\hbar g_J \mu_{\rm B} \hat{\bm{J}} \cdot \bm{B}$ is the Zeeman hamiltonian.



In general, the light shift hamiltonian does not commute with the Zeeman hamiltonian. We work here in a regime where the Zeeman shift 
($\sim\mu_{\rm B}\,B\,g_J m_J>\SI{40}{\mega\hertz}$ for $|m_J|=8$ $B>\SI{3}{\gauss}$) is much larger than the vector or tensor light shift ($\sim\alpha_{V} \frac{I}{2c \epsilon_0}m_J/J$ or $\sim\alpha_{T} \frac{I}{2c \epsilon_0}m_J^2/J^2$, a few MHz at most), such that the eingenstates of the total hamiltonian are always close to the Zeeman states \footnote{We also performed full diagonalization of the hamiltonian a posteriori to verify that this assumption is correct.}. In these conditions, the effect of the light shift hamiltonian can be described simply as a first-order energy shift of the Zeeman states and the trapping potential created by a dipole trap follows $V(\bm{r}) = -\alpha\frac{I(\bm{r})}{2c\epsilon_0}$  with a polarizability $\alpha$ that depends on the tweezer polarization and the Zeeman state $m_J$.
In particular, one finds that atoms in the most negative Zeeman state $m_J=-J$ experience an effective polarizability~:
\begin{equation}
    \alpha = \as -i \frac{\av}{2} (\bm{\epsilon}^*\times\bm{\epsilon})\cdot\bm b + \frac{\at}{2}\left[3(\bm{\epsilon}\cdot\bm b)(\bm{\epsilon}^*\cdot\bm b)-1\right]
\end{equation}
where $\bm b=\bm{B}/\lVert \bm{B}\rVert$ is the direction of the magnetic field.

Since the direction of the magnetic field plays an important role in our measurement, we carefully calibrated the applied magnetic fields using Zeeman spectroscopy in order to compensate for stray fields.
Below we report measurements of the scalar, vector and tensor polarizabilities for both the ground state $G$ and the upper state $E$ of the intercombination line in a trap with a wavelength of \SI{532}{\nano\meter} for $^{162}$Dy.
To do so, we first perform experiments in a dipole trap containing many atoms. The atoms are imaged in time-of-flight so that the magic condition is not required and we can vary the trap polarization. 
We increase the size of the trap by reducing the input pupil of the microscope with an iris, from an initial NA of 0.5 to an effective NA of about 0.2. 
In this case, the trap intensity can be modeled by an Airy pattern for which the radial and axial intensity profiles near the center of the trap are respectively given by:
\begin{equation}
\label{eq:radial-intensity}
    I(r, z=0)  =  I_0  \left[\frac{2J_1(k\,r\,\NA)}{k\,r\,\NA}\right]^2 
\end{equation}
and 
\begin{equation}
\label{eq:axial-intensity}
    I(r=0, z) =  I_0\,\mathrm{sinc}^2\left(\frac{k\,z\,\NA^2}
    {4}\right)
\end{equation}
where $k=2\pi/\lambda_{\rm trap}$ is the wavevector for the trapping light, $I_0=Pk^2\,\NA^2 / 4\pi$ is the peak intensity, $P$ is the trap power and $J_1$ is the Bessel function of the first kind with order 1.
We found that this intensity profile matches better what was observed on the experiment than what a Gaussian profile predicts. In a test setup reproducing the experimental conditions, we measured the intensity profile and we verified that it matches an Airy profile.

\subsection{Measurement of polarizabilities of the ground state $G$}\label{sec:gnd-polar}

To estimate the ground state polarizabilities, we measure the mechanical oscillation frequencies of the dipole trap for atoms in the state $|g\rangle = |G, J=8, m_J=-8\rangle$ at different trap polarizations. 
The frequencies of the trap are obtained by approximating the potential experienced by the atom with a harmonic potential $V(\bm{r}) \simeq -U_0 + \frac{1}{2}m\omega_\perp ^2(x^2 + y^2) + \frac{1}{2}m\omega_\parallel ^2 z^2$.
In this equation, $m$ is the atomic mass, $\omega_\perp = k\,\NA\sqrt{\frac{U_0}{2m}}$ and $\omega_{\parallel} =k\,\NA^2\sqrt{\frac{U_0}{24m}}$ are respectively the radial and axial trap frequencies and $U_0=\frac{\alpha I_0}{2c\epsilon_0}$ is the trap depth.

\begin{figure}
    \includegraphics[width=\columnwidth]{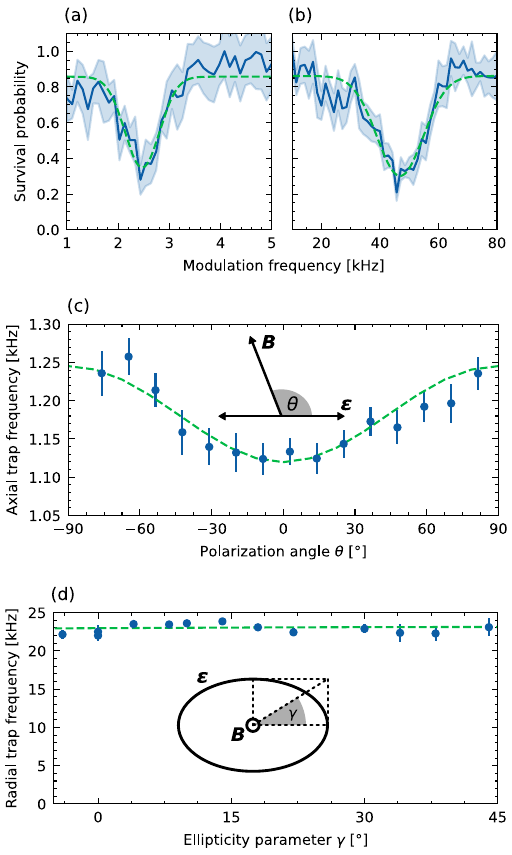}
    \caption{Measurement of the axial (a) and radial (b) oscillation frequencies in the ground state using parametric heating for a tweezers power $P=\SI{14}{\milli\watt}$.
    The dashed lines are fits to Gaussian dips in order to locate the resonant modulation frequency.
    We measure significant losses when we modulate the trap intensity at $2.47(7\sys)(2\stat)$~\unit{\kilo\hertz} and $47(5\sys)(1\stat)$~\unit{\kilo\hertz}, corresponding respectively to twice the axial frequency and twice the radial frequency.
    (c) Axial trap frequency for atoms in the state $|G, J, m_J=-J\rangle$ as a function of the angle $\theta$ between the trap linear polarisation $\bm{\epsilon}$ and the quantization axis defined by the magnetic field $\bm{B}$.
    Blue dots are experimental points.
    The dashed green line is a fit to equation (\ref{eq:linear-frequency}).
    (d) Radial trap frequency in the state $|G, J, m_J=-J\rangle$ as a function of the ellipticity parameter $\gamma$.
    Blue dots are experimental points.
    The dashed green line is a fit to equation (\ref{eq:elliptic-frequency}) and appears flat due to the negligible vector polarizability in the ground state.
    }\label{fig:trap-frequencies}
\end{figure}

To measure the trap frequencies, we use parametric heating by modulating the trap intensity by \SI{10}{\percent} of its nominal value for \SI{50}{\milli\second}, at variable frequencies. 
A typical measurement of the remaining atom number after modulation as a function of modulation frequency is shown in figure~\ref{fig:trap-frequencies} showing dips at twice the axial (a) and radial (b) frequency of the trap. 

The power in the tweezer for the measurements presented in this part is $P=14(3\sys)$~\unit{\milli\watt}.
The uncertainty on experimental values is written in parentheses behind the measured value.
In this case, it is a systematic uncertainty coming from the estimated transmission of the beam path.
In addition, the fact that the atoms are not trapped precisely at the trap bottom, combined with the fact that the trap is not perfectly harmonic, creates a systematic error in the obtained frequency \cite{Ravensbergen2018}, we estimate a 10\,\%  systematic error in our experiments. 

The ratio of frequencies yields the effective numerical aperture of the trap $\NA = 2\sqrt{3} \omega_\parallel / \omega_\perp = 0.182(20\sys)(6\stat)$ in agreement with our expectation. 
Here the two values in parentheses refer respectively to the systematic and statistical uncertainties.
To analyze the data below, we use this experimental value to infer the trap intensity using equations\,(\ref{eq:radial-intensity},\,\ref{eq:axial-intensity}).

To extract the polarizabilities of the ground state, we first perform experiments with a linear trap polarisation $\bm{\epsilon}$. In this case, the vector contribution to the light-shift vanishes and the potential experienced by an atom in $m_J=-8$ is:
\begin{equation}
V^{\rm lin}(\bm{r})=-\frac{I(\bm{r})}{2c\epsilon_0}\left(\as^G + \frac{\at^G}{2}[3\cos^2(\theta)-1]\right)
\end{equation}
where $\theta$ is the angle between the polarization of the trapping light and the magnetic field $\bm{B}$.
We change the value of the parameter $\theta$ by rotating the magnetic field
 with respect to a fixed tweezer polarization.
 For each angle we measure the axial trap frequency (see figure~\ref{fig:trap-frequencies}(c)) and we fit the obtained data to:
\begin{equation}\label{eq:linear-frequency}
 \omega_{\parallel}(\theta) = \omega_{\parallel, 0}\sqrt{1+\frac{\at^G}{2\as^G}[3\cos^2(\theta)-1]}
\end{equation}
From this analysis, we find $\omega_{\parallel, 0}=2\pi\times 1.21(10\sys)(2\stat)$~\unit{\kilo\hertz}. 
This yields $\alpha^G_S=150(100\sys)(30\stat)\times \alpha_0$, with $\alpha_0=\SI{1.649e-41}{\coulomb^2\meter^2\joule^{-1}}$ being the atomic unit of polarizability.
Similarly, the fitted curve gives us $\at^G / \as^G = -0.14(6\stat)$.
It is worth noting that there is a large systematic error for the absolute polarizabilities due to the uncertainty on the light intensity at the position of the atoms, but this unknown does not affect the ratio of the two polarizabilities.\newline

To measure the vector polarizability $\av^G$, the trap polarization must have a circular component.
We use an elliptical polarization in the plane perpendicular to the magnetic field $\bm{B}=B\bm{e}_z$ aligned to the tweezer axis:
\[\bm{\epsilon} = \cos(\gamma)\bm{e}_x + i \sin(\gamma)\bm{e}_y\]
By scanning the parameter $\gamma$ from 0\textdegree\! to 45\textdegree\!, the polarization can be changed from linear to circular.
The potential experienced by atoms in the $|G, J, m_J=-J\rangle$ state is:
\begin{equation}
V^{\rm el}(\bm{r})=-\frac{I(\bm{r})}{2c\epsilon_0}\left(\as^G -\frac{\at^G}{2} + \frac{\av^G}{2}\sin(2\gamma)\right)
\end{equation}
We modify this ellipticity parameter by sending the tweezer beam through a quarter waveplate with variable angle. We repeat the measurement of the trap frequencies for different ellipticity parameters $\gamma$ by rotating the quarter-wave plate and fit the resulting radial trap frequencies (figure~\ref{fig:trap-frequencies}(d)) to:
\begin{equation}\label{eq:elliptic-frequency}
    \omega_\perp (\gamma) = \omega_{\perp, 0} \sqrt{1 + \frac{\av^G}{2\as^G-\at^G}\sin(2\gamma)}
\end{equation}
As visible in figure \ref{fig:trap-frequencies}(d), the dependence of the trap frequency on $\gamma$ is smaller than experimental uncertainties, yielding a very small vector polarizability in the ground state: $\frac{\av^G}{2\as^G-\at^G}=0.01(4\stat)$.

\subsection{Differential polarizabilities on the intercombination transition}\label{sec:diff-polar}

\begin{figure}
    \includegraphics[width=\columnwidth]{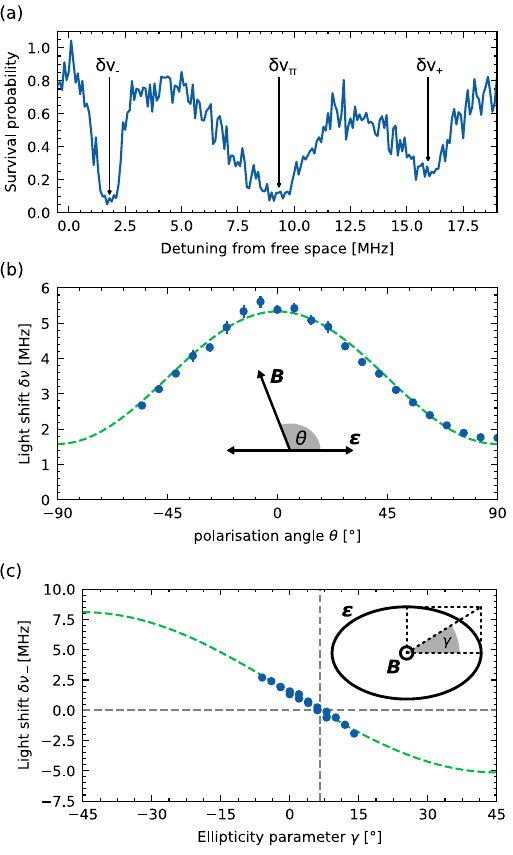}
    \caption{(a) Induced losses in the dipole trap when shining \SI{626}{\nano\meter} light as a function of light frequency. The effective numerical aperture used here is $NA = 0.18$ and the tweezer power is $P=\qty{33}{\milli\watt}$. 
    (b) Differential light shift between $|G, J, m_J=-J\rangle$ and $|E, J', m'_{J}=-J'\rangle$ for a linear polarization as a function of $\theta$. 
    (c) Differential light shift between $|g\rangle$ and $|e\rangle$ for an elliptical polarization as a function of $\gamma$.
    Blue dots are experimental points.
    The dashed green line is a fit to equation (\ref{eq:elliptic-light-shift}).
    }
    \label{fig:resonant-losses}
\end{figure}

Once the polarizabilities of the ground state $G$ have been measured, we use these values to deduce the polarizabilities of the excited state $E$.
To do this, we measure the differential light shift on the $|g\rangle = |G,~J=8,~m_J=-8\rangle \leftrightarrow |e\rangle = |E,~J'=9,~m_J'=-9\rangle$ transition in the trap.
When shining \SI{626}{\nano\meter} light on many atoms in the trap, losses are induced if the light frequency is resonant with a transition between the ground state and the excited state.
Figure~\ref{fig:resonant-losses} shows an example of such experiment.
In this case, we apply a magnetic field of \SI{3}{\gauss} to separate the $\sigma_+$, $\sigma_-$ and $\pi$ transitions.
We then shine \SI{626}{\nano\meter} light at $I\simeq 5\, I_{\rm sat, 626}$ for \SI{20}{\milli\second} and observe atomic losses when the frequency of this light is resonant with a transition between Zeeman levels of the ground states and excited states.
The effective $\NA=0.18$ used for these measurements is the same as above and the power of the dipole trap beam is $P=33(6\sys)$~\unit{\milli\watt}.
The frequency is referenced from the frequency of the transition $|g\rangle \leftrightarrow |e\rangle$ in free space.
There are three dips in the signal, corresponding to the three projections of the \SI{626}{\nano\meter} light polarization onto the magnetic field axis.
Since the atoms are pumped in the $|g\rangle$ state in the MOT before being loaded in the trap, only the $\sigma_-$ transition is cycling.
Atoms interacting with $\sigma_+$ and $\pi$ light are pumped towards other $m_J$ states, causing a broadening of these two transitions. We therefore rely only on the frequency of the $\sigma^-$ transition to extract the differential polarizability.

We measured the differential light-shift again for different polarizations of the trap beam. In figure~\ref{fig:resonant-losses}(b), we rotate the magnetic field direction with respect to a linear trap polarization to measure the differential scalar and tensor polarizabilities $\Delta\as=\as^\text E-\as^\text G$, $\Delta\at=\at^\text E-\at^\text G$.
We fit this curve to:
\begin{equation}
    \begin{split}
        \delta\nu(\theta) & = \frac{V_E^{\rm lin}(0)-V_G^{\rm lin}(0)}{h}\\
        &= \delta\nu_0 \left[1+\frac{\Delta\at}{2\Delta\as}\left(3\cos^2(\theta)-1\right)\right]
        \end{split}
\end{equation}
and find $\delta\nu_0=-\Delta\as I_0 / 2 h c\epsilon_0=2.83(3\stat)$~\unit{\mega\hertz} and $\frac{\Delta\at}{\Delta\as}=0.86(3\stat)$.
Assuming that the intensity is given by equations~\eqref{eq:radial-intensity} and~\eqref{eq:axial-intensity}, this gives $\Delta\as = -50(14\sys)(3\stat)\times\alpha_0$.\newline

To measure the differential vector polarizabilities $\Delta\av=\av^\text E-\av^\text G$, we perform a similar measurement, now changing the ellipticity parameter of the trap polarization.
This is again done in a configuration where the polarization ellipse is perpendicular to the magnetic field (see figure~\ref{fig:resonant-losses}(c)).
These values are fitted to:
\begin{equation}\label{eq:elliptic-light-shift}
    \begin{split}
        \delta\nu (\gamma)& = \frac{V_E^{\rm el}(0)-V_G^{\rm el}(0)}{h} \\
                  & = \delta\nu_0 \left[1+\frac{\Delta\av}{2\Delta\as-\Delta\at}\sin(2\gamma)\right]
    \end{split}
\end{equation}
with $\delta\nu_0=-(\Delta\as-\Delta\at/2) I_0 / 2 h c\epsilon_0=1.48(5\stat)$~\unit{\mega\hertz} and $\Delta\av / (2\Delta\as-\Delta\at)=-4.46(1\stat)$.\newline

An important conclusion from our measurements is that there is a specific ellipticity parameter $\gamma^*=6.5(2\stat)$\unit{\degree} (vertical dashed line in figure \ref{fig:resonant-losses}(c)) such that the differential light shift cancels for the two states $|g\rangle$ and $|e\rangle$.
This is a so-called magic ellipticity and in this configuration, the difference of energy between the two states is independent of the trap intensity.
This condition proves particularly useful to realise robust probing or imaging of Dy atoms in a deep trap, as discussed earlier.

\subsection{Polarizability measurement free from trap waist errors}\label{sec:move}

The absolute value of the scalar polarizability that we extracted above is far from theoretical expectations (see discussion below) and is accompanied by a large uncertainty due to the difficulty to calibrate the trap shape in-situ.

To get a more accurate value, we perform an in-situ measurement of two observables, the trap frequency and the trap depth which yields a measurement of the polarizability that is \emph{independent of the trap size}, as was performed for instance using another species as a reference in \cite{Ravensbergen2018}. This method is valid for any trap intensity profile whose radial dependence is defined by a single length scale: $I(r)= \frac{P}{2\,\pi\,l_r^2}\,f(r/l_r)$. For Gaussian beams or diffraction-limited beams \eqref{eq:radial-intensity} as is the case here, the following relationship holds \footnote{Equation \eqref{eq:UOm} holds assuming $\zeta=-\frac{f(0)^2}{f''(0)}=1$, this is true for Gaussian or diffraction-limited beams, but for higher-order ones such as Laguerre-Gauss, one then needs to multiply the right hand side of \eqref{eq:UOm} by $\zeta$}:  
\begin{equation}
\left(\frac{U_0}{\omega_\perp}\right) ^ 2 = \frac{\alpha m P}{4\pi c \epsilon_0}\;\label{eq:UOm}
\end{equation}
with $U_0$ the potential depth of a tweezer in the ground state, $\omega_\perp$ the radial trapping frequency, and $P$ the beam power.  
One then only needs to measure $P$, $\omega_\perp$ and $U_0$ for the same trap to extract $\alpha$. 
While the two former are straightforward, we use here a method based on single atoms in moving tweezers to measure $U_0$.


\begin{figure}
    \centering
    \includegraphics{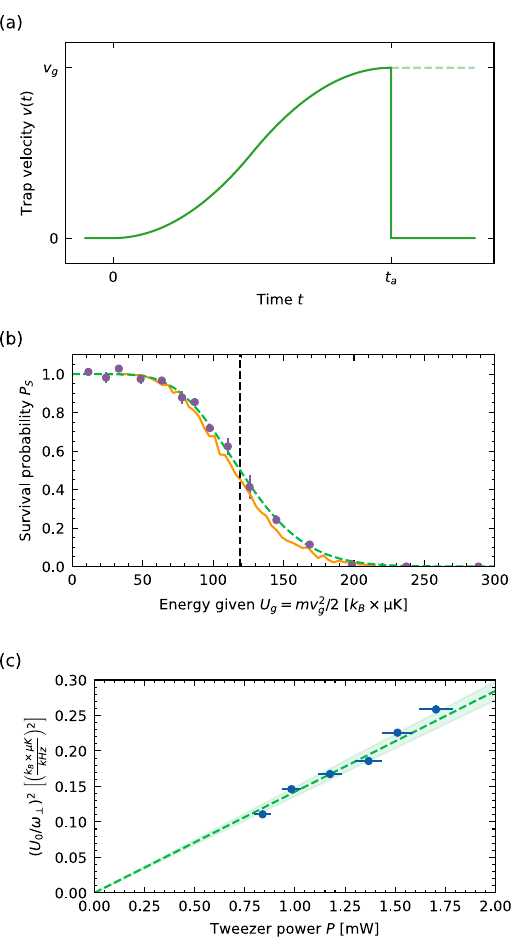}
    \caption{
    (a) Trajectory followed by the trap to measure the trap depth.
    The trap is adiabatically accelerated to a velocity $v_g$ and then abruptly brought to a stop.
    (b) Fraction of the atoms that survives the abrupt stop of the traps initially moving at the velocity $v_g$.
    Purple points are experimental measurements. 
    They are corrected to take into account losses during imaging. 
    The green line is a fit using equation (\ref{eq:trap-depth}). The orange solid line shows results of 3D simulations of atoms in a moving Gaussian trap.
    The dashed vertical line indicates the trap depth.
    (c) Evolution of $(U_0 / \omega_\perp)^2$ for different tweezer powers.
    Blue dots are experimental points.
    Dashed green line is linear fit to the data.
    }
    \label{fig:trap-depth}
\end{figure} 

To do so, we rely on the fact that the single-atom tweezers' position is controllable in real time by the acousto-optic deflector that produces the traps. We perform a sequence where we place an atom in a tweezer, that we adiabatically accelerate to a given velocity $v_g$ in the direction perpendicular to the chain.
In a tweezer, the atom is at rest in the moving frame of the tweezer, but reaches the velocity $v_g$ in the lab frame.
We then abruptly stop the tweezer.
The atom now has a kinetic energy $E_K=\frac{1}{2}m v_g^2$ in the trap frame.
If $E_K$ is smaller than the trap depth $U_0$, the atom will remain in the trap.
However, if $E_K$ is larger than $U_0$, the atom will be expelled from the trap.
The exact trajectory followed by the traps is shown on figure \ref{fig:trap-depth}(a).
At time $t=0$, the trap is slowly accelerated until it reaches the constant velocity $v_g$ where it is then abruptly stopped.
Typically the distance covered by the trap in this time is $\Delta x = 43(1\sys)$~\unit{\micro\meter} and we change the velocity $v_g$ reached at the end of the motion by scanning the acceleration time $t_a$ between \SI{1}{\milli\second} and \SI{5}{\milli\second}.
One can then look for the threshold in velocity $v_g$ given to the atoms under which they remain in the trap (see blue curve in fig.~\ref{fig:trap-depth}(b)).
       
In practice, the threshold is broadened by the non-zero atomic temperature.
The initial velocity $v$ of the atom along the direction of motion follows the Boltzman distribution:
\[
P(v) = \sqrt{\frac{m}{2\pi k_B T}}\exp\left(-\frac{1}{2}\frac{mv^2}{k_B T}\right)
\]
After the trap has stopped, the atom then has the kinetic energy $E_K(v) = \frac{1}{2}m(v+v_g)^2$.
The probability that the atom remains in the trap is then:
\begin{equation}\label{eq:trap-depth}
\begin{split}
    P_S & =  \int  \Theta\left(U_0 - E_K(v)\right)P(v)\mathrm{d}v \\
    & = 
\frac{1}{2}\left[
\mathrm{erf}\left(\sqrt{\frac{U_0}{k_B T}} + \sqrt{\frac{U_g}{k_B T}}\right)
+\mathrm{erf}\left(\sqrt{\frac{U_0}{k_B T}} - \sqrt{\frac{U_g}{k_B T}}\right)
\right]
\end{split}
\end{equation}
where $U_g = \frac{1}{2}m v_g^2$ is the kinetic energy imparted by the tweezer movement.
Thus, under the assumption of adiabaticity, the result of this experiment only depends on the trap depth and atomic temperature, but does not depend on the exact trap shape.


In figure \ref{fig:trap-depth}(b), we show the result of experiments performed in parallel on 50 tweezers whose powers have been equalized \cite{Bloch2023}.  
We ensure that the polarization is homogeneous across the traps by placing a Glan-Thompson polarizer directly after the AOD.
Atoms are imaged once before the movement and once after. 
Averaging over the 50 tweezers and many realizations yields the recapture probability $P_S$ for a given velocity $v_g$. 
The solid green line is a fit with equation \eqref{eq:trap-depth}, which yields a trap depth of $U_0=k_B\times 119(7\sys)(1\stat)$~\unit{\micro\kelvin} and a temperature $T=5.5(4\stat)$~\unit{\micro\kelvin} for a tweezer power of $1.7(3\sys)$~\unit{\milli\watt}. 
We find that this temperature is in excellent agreement with the usual release-recapture method \cite{Bloch2023}. 
We have verified that numerical simulations of the Newton equations of motion for atoms in a moving Gaussian trap reproduce very well the results (orange solid line in figure \ref{fig:trap-depth}(b)).
In figure \ref{fig:trap-depth}(c), we show the value of $(U_0/\omega_\perp)^2$ obtained for variable tweezer power, exhibiting a linear dependence.
A linear fit gives the atomic polarizability of $\alpha=204(40\sys)(5\stat)\times \alpha_0$. This measurement was realised on single atoms, in the magic imaging configuration with a polarization ellipse perpendicular to the magnetic field and with ellipticity parameter $\gamma^*=6.5$.
In this configuration $\alpha=\as^G -\at^G/2 + \av^G/2\sin(2\gamma^*)$. Using the values of the atomic polarizability ratios from section \ref{sec:gnd-polar}, we obtain a better estimate of the ground state scalar polarizability $\as^G=180(35\sys)(10\stat) \times \alpha_0$. 
This value is compatible with the one obtained in the previous section ($\as^G=150(100\sys)(30\stat) \times \alpha_0$), in a different trap, ruling out a large systematic error not taken into account. 

\section{Comparison with calculations}

We use the more accurate estimation of the absolute value of $\as^G$ obtained in section \ref{sec:move} in addition to the ratios measured in sections \ref{section:polarizabilities}A, B to calculate all the polarizabilities in the ground state and excited state.
They are summarized in table \ref{tab:polarizabilities}, left and on figure 7. 
We discuss below how they compare with theoretical expectations.

\begin{table}[h]
  \centering
    \begin{tabular}{| c || c | c |}
    \hline
        \multicolumn{3}{|c|}{Experiment}\\
        \hline
        \hline
        & $G$ &  $E$ \\
        \hline
        \hline
        $\as$ & \SI{180(36)}{} & \SI{130(40)}{}\\
        \hline
        $\av$ & \SI{4(15)}{} & \SI{260(80)}{}\\
        \hline
        $\at$ & -\SI{25(12)}{} & -\SI{68(18)}{}\\
        \hline
    \end{tabular}
    \hspace{10pt}
    \begin{tabular}{| c || c | c |}
        \hline
        \multicolumn{3}{|c|}{Theory}\\
        \hline
        \hline
        & $G$ &  $E$ \\
        \hline
        \hline
        $\as$ & \SI{408(56)}{} & \SI{73(30)}{}\\
        \hline
        $\av$ & -\SI{57(93)}{} & \SI{125(60)}{}\\
        \hline
        $\at$ & -\SI{18(61)}{} & -\SI{44(38)}{}\\
        \hline
    \end{tabular}
    \caption{Left: Experimentally measured polarizabilities for the ground state and the upper state of the intercombination line at \SI{532}{\nano\meter}, in atomic unit ($\alpha_0=\SI{1.649e-41}{\coulomb^2\meter^2\joule^{-1}}$). 
    Values in parentheses are the quadratic sum of systematic and statistical uncertainties.
    Right: Theoretical values and their estimated uncertainties.}
    \label{tab:polarizabilities}
\end{table}

\begin{figure}
    \centering
    \includegraphics{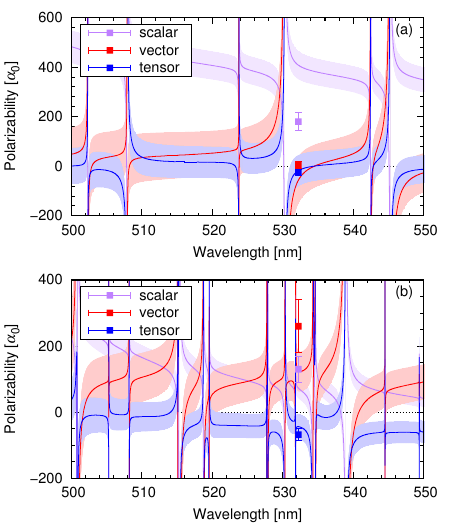}
    \caption{
    Calculated scalar (purple), vector (red) and tensor (blue) polarizabilities, as well as their uncertainties, as functions of the trap wavelength:
    (a) of the level $|G\rangle$;
    (b) of the level $|E\rangle$.
    Experimental values at 532.208\,nm are also shown with their error bars.
    }
    \label{fig:pola-wvl}
\end{figure}





We report in table \ref{tab:polarizabilities}, right, and in figure \ref{fig:pola-wvl} the theoretical expectations for the polarizabilities and their uncertainties obtained by using the methods presented in \cite{Li2017, Chalopin2018a} and modified in such a way that, when they are known, experimental level energies are included in a sum-over-states formulas. We consider that the uncertainty on polarizabilities comes from the uncertainty on theoretical line strengths, for which Ref.~\cite{Li2017} gives an average value of 12.4~\%. Each term of the sum-over-state formulas brings to the uncertainty a contribution equal to its absolute value multiplied by 12.4~\%. As one can see, despite the rather large  uncertainties, the experimental polarizabilities that we obtained do not all agree with theoretical expectations.

In particular, the disagreement is sizeable for the scalar polarizability, with the experiment a factor 2 below expectations. The measured ground state polarizability is close to that measured at 1064\,nm ($\alpha_{\text S}=\SI{184.4(2.4)}{}\,\alpha_0$, in agreement with calculations at that weavelength performed with the same method) \cite{Ravensbergen2018}. Since the wavelength of 532\,nm is closer to the strong blue transitions of Dy, which are mainly contributing to the scalar polarizability $\as$, one would expect $\as$ to be higher at this wavelength than at 1064\,nm, as predicted in \cite{Li2017}. A lower value could be explained if there was a nearby transition from which the trap wavelength (precisely 532.208\,nm) would be blue-detuned. The closest candidate would be the level at 18528.55~cm$^{-1}$ (corresponding to a transition wavelength of 539.708~nm). But its transition with the ground state has not been detected \cite{Wickliffe2000}, and our calculations predict an Einstein coefficient $A_{ik} \approx 7 \times 10^4$~s$^{-1}$, much too weak to affect $\as$ at 532~nm. On the contrary, our trap wavelength is red-detuned with respect to the level at 18857.04~cm$^{-1}$ (530.306\,nm), which tends to increase the scalar polarizability. Moreover, our theoretical calculations including the configuration [Xe]4f$^{9}$5d$^{2}$6s do not predict any experimentally unknown level in this region of the Dy spectrum.

Regarding the other values, we note that the vector and tensor polarizabilities in the ground state and excited state do not perfectly agree but fall within the ranges of uncertainty. However, since these value are extracted from the knowledge of the ground scalar polarizability, the mismatch would be higher if it was wrong by a factor 2. Since the experimental methods used here are free of errors on the trapping laser intensity, and consistent over several measurements, the origin of the theory-experiment disagreement remains yet unknown. The present work thus begs for future investigations to understand this large discrepancy. \\

\section{Conclusion}

To conclude, we have presented a new experimental apparatus  where a MOT of Dy in a glass cell is continuously loaded from a 2D MOT. This MOT allows to load optical tweezers at 532 nm. We reported here measurements of all components of the polarizability on the intercombination transition. We find a remarkably low scalar polarizability in the ground state, which calls for future investigations. Our setup allows us to image single atoms in magic tweezers with $>99\,\%$ fidelity and about $2.5\,\%$ loss probability in a single image, allowing for many future studies using the tweezers toolbox.

To improve imaging fidelity when using the intercombination line, we expect that it would be beneficial to use optical tweezers with a longer wavelength, which would suppress excitation of atoms to high-lying metastable states. One could use for instance the magic condition found in \cite{Chalopin2018a}. Finally, in this work we used exclusively the bosonic $^{162}$Dy isotope which has no hyperfine structure.
However, we expect that fermionic isotopes with their hyperfine structure might present different and possibly advantageous mechanisms for imaging and deterministic loading of the tweezers, as was observed for example on ytterbium \cite{Jenkins2022}.

\begin{acknowledgments}
We thank Florence Nogrette for experimental help and discussions. We note that another setup based on a similar 421-nm 2D-MOT loading a 626-nm 3D-MOT of Dy atoms has been developed in the group of L. Chomaz \cite{jin2023}. We have widely benefited from exchanges between our groups. This project has received funding by the Agence Nationale de la Recherche (JCJC grant DEAR, ANR-22-PETQ-0004 France 2030, project QuBitAF), by the European Union (ERC StG CORSAIR,  101039361, ERC AdG ATARAXIA 101018511), and the Horizon
Europe programme HORIZON-CL4- 2022-QUANTUM-02-
SGA (project 101113690 (PASQuanS2.1)).
\end{acknowledgments}

\bibliography{biblio}



\end{document}